\documentclass{moriond}
\usepackage{amssymb,amsmath}
\def\be{\begin{equation}}
\def\ee{\end{equation}}
\def\bea{\begin{eqnarray}}
\def\eea{\end{eqnarray}}

\newcommand{\Photo}{}

\begin{document}
\begin{flushright}
LAPTH-041/14
\end{flushright}
\vspace*{3.5cm}
\title{EVALUATING THE 6-POINT REMAINDER FUNCTION\\
 NEAR THE COLLINEAR LIMIT}

\author{G.~PAPATHANASIOU}

\address{LAPTh, CNRS, Universit\'e de Savoie, Annecy-le-Vieux F-74941, France
\vskip0.20cm
\textup{and}
\vskip0.20cm
Physics Department, Theory Division, CERN, CH-1211 Geneva 23, Switzerland}

\maketitle\abstracts{The simplicity of maximally supersymmetric Yang-Mills theory makes it an ideal theoretical laboratory for developing computational tools, which eventually find their way to QCD applications. In this contribution, we continue the investigation of a recent proposal by  Basso, Sever and Vieira, for the nonperturbative description of its planar scattering amplitudes, as an expansion around collinear kinematics. The method of \texttt{arXiv:1310.5735}, for computing the integrals the latter proposal predicts for the leading term in the expansion of the 6-point remainder function, is extended to one of the subleading terms. In particular, we focus on the contribution of the 2-gluon bound state in the dual flux tube picture, proving its general form at any order in the coupling, and providing explicit expressions up to 6 loops. These are included in the ancillary file accompanying the version of this article on the \texttt{arXiv}.}

\section{Introduction and Summary}

Maximally supersymmetric Yang-Mills theory (MSYM) offers a unique possibility for the nonperturbative investigation of gauge theories. In its strongly coupled regime it can be mapped to weakly coupled strings of type IIB on $AdS_5\times S^5$, which are amenable to perturbative computations. Furthermore, in the planar limit, where the number of colors $N$ goes to infinity with the 't Hooft coupling $\lambda\equiv g^2_{YM}N$ fixed, integrable structures emerge, which allow the determination of certain quantities to all loops~\cite{Beisert:2010jr}. More importantly, by being the simplest 4-dimensional interacting gauge theory, it serves as an excellent theoretical laboratory for developing computational tools, before applying them to QCD. Celebrated examples of this strategy are generalized unitarity for scattering amplitudes and more recently the method of symbols, for an overview see~\cite{Dixon:2011xs} and references therein. The symbol has been used in calculations of several QCD processes, such as gluon fusion to heavy quark-antiquark pair~\cite{Bonciani:2013ywa}, relevant to experiments at the LHC.



\begin{figure}
\begin{minipage}{0.5\linewidth}
\centerline{\includegraphics[width=0.6\linewidth]{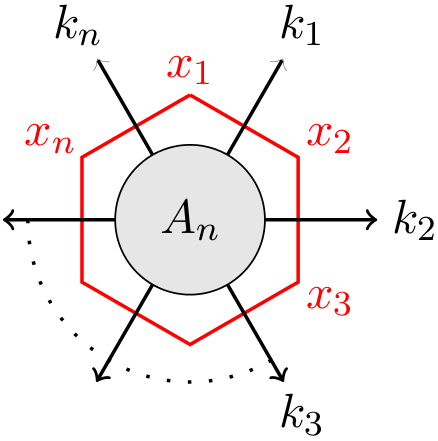}}
\end{minipage}
\hfill
\begin{minipage}{0.49\linewidth}
\centerline{\includegraphics[width=0.7\linewidth]{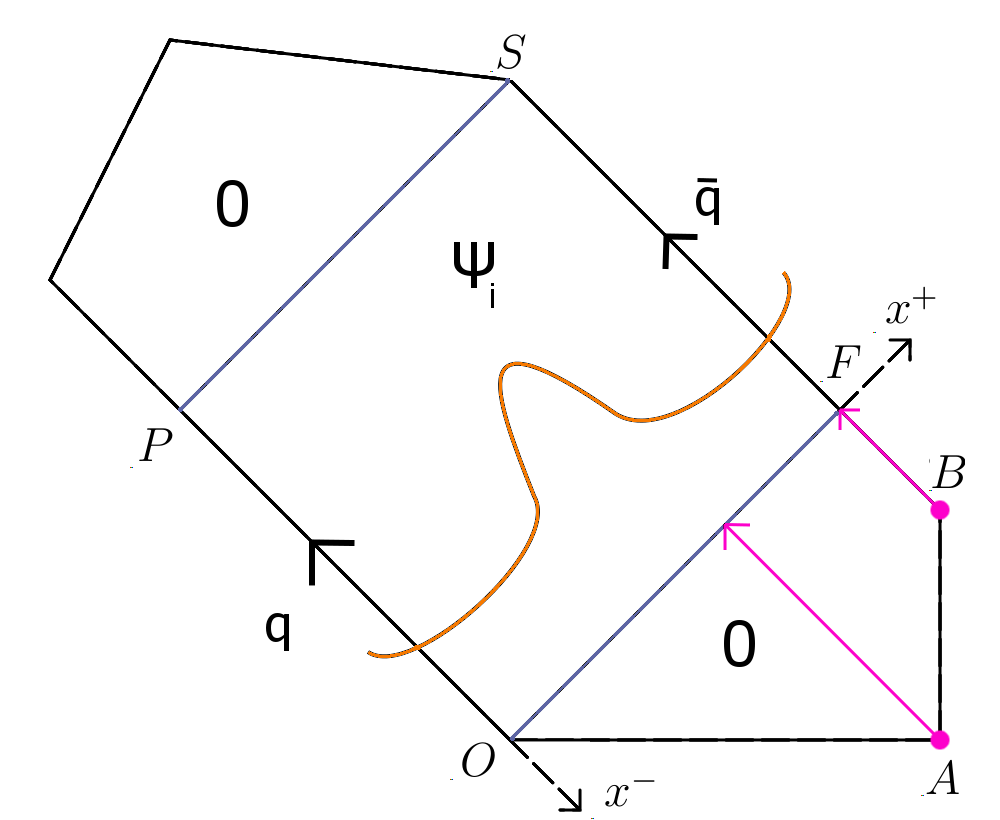}}
\end{minipage}
\caption[]{Left: The $n$-point MHV amplitude $A_n$ may be identified with a Wilson loop $W_n$ after defining dual space variables $k_i\equiv \textcolor{red}{x_{i+1}}-\textcolor{red}{x_i}\equiv x_{i+1,i}$. Since $k_i^2=0$, all distances between the cusps $x_i$ will be lightlike. Right: To take the collinear limit of $W_6$, we first connect two non-adjacent edges and form a square $(OPSF)$. It is invariant under three symmetries, and we act with one of them on $A$ and $B$, to flatten them on $(OF)$. We can then think of $(PO),(SF)$ as a color-electric flux tube sourced by $q\bar q$, and decompose $W_6$ w.r.t. its excitations $\psi_i$.}
\label{fig:WLduality}
\end{figure}

In this contribution, we will focus on the near-collinear kinematics of the planar, Maximally Helicity Violating (MHV) 6-point amplitude of MSYM. Planarity has the benefit that the only surviving color structure is a single trace of generators in the adjoint representation of the gauge group, which we can strip off in order to study its coefficient, the color-ordered amplitude. Among all different helicity configurations for the external gluons of such amplitudes, it turns out that the MHV ones $A(+\cdots+--)$, corresponding to all but two helicities being the same, are the simplest. Remarkably, these amplitudes have been also observed to be dual to Wilson loops made of straight lightlike segments, as shown in figure \ref{fig:WLduality}. And the fact that for $n=4,5$ legs the dimensionally regulated amplitude is accurately described to all loops by the ansatz of Anastasiou-Bern-Dixon-Kosower/Bern-Dixon-Smirnov, implies that the 6-point amplitude is indeed the next interesting case to consider. The \emph{remainder function} is precisely the part of the amplitude not captured by the aforementioned ansatz. An account of these developments may be found in~\cite{Dixon:2011xs} as well.

Last but not least, there is growing evidence that each term in the expansion of the remainder function around the limit where consecutive external momenta become collinear, can be computed to all loops with the help of integrability, see~\cite{Basso:2013vsa,Basso:2013aha,Basso:2014koa} and references therein. 
For the $l$-loop 6-point remainder function $R_6^{(l)}$, symmetry implies that this expansion has the form
\be\label{R6_general}
R_6^{(l)}=\sum_{m=1}^\infty e^{-m\tau}\sum_{p=0}^{[m/2]} \cos [(m-2p)\phi]\sum_{n=0}^{l-1} \tau^n f_{m,p,n}^{(l)}(\sigma)\,,
\ee
where $[x]$ denotes the integer part of $x$, and $\{\tau,\phi,\sigma\}$ a convenient choice of kinematical variables, in which the collinear limit is described by $\tau\to\infty$. As we illustrate in figure \ref{fig:WLduality}, each term in the sum in $m$ receives contributions from all $m$-particle excitations of a color-electric flux tube, created by the two segments adjacent to the ones becoming collinear, whose dynamics are encoded in an integrable spin chain.

These excitations may also be thought of as insertions the fields of the theory on the side $(OF)$ of the middle square in figure \ref{fig:WLduality}. The leading $m=1$ term in (\ref{R6_general}) comes from a single gluon insertion, and integral expressions for it were found~\cite{Basso:2013vsa,Basso:2013aha} by exploiting the aforementioned integrable structures. In~\cite{Papathanasiou:2013uoa}, we analyzed these integrals, and proved that at any loop order,
\be
f^{(l)}_{1,0,n}(\sigma)=\sum_{s,r,m_i} c^{\pm}_{s,m_1,\ldots,m_r} e^{\pm\sigma}\sigma^s H_{m_1,\ldots,m_r}(-e^{-2\sigma})\,,\quad m_i\ge1\,,
\ee
where $c^{\pm}$ are numeric coefficients, and $H_{m_1\ldots m_n}$ transcendental functions known as harmonic polylogarithms (HPLs). Our proof constituted an algorithm for the direct evaluation of the integrals for arbitrary $l$, which we employed in order to obtain $f_{1,0,n}^{(l)}$ for any $n$ up to $l=6$ loops, and for $n=l-1$ up to $l=12$ loops.

More recently, the $m=2$ particle excitations were analyzed, and all-loop integral expressions were also presented for the corresponding term in (\ref{R6_general})~\cite{Basso:2014koa}. A variety of different flux tube excitations contributes in this case, and here we will focus on the 2-gluon bound state $DF$, whose contribution $\mathcal{W}_{DF}$ is part of $f_{2,0,n}^{(l)}$. Extending the method of~\cite{Papathanasiou:2013uoa}, we similarly prove that~\footnote{Note that our definitions for the $\mathcal{W}$'s  here and in eq.~(\ref{calW_excitations}) differ from those of~\cite{Basso:2014koa}, in that we have taken out the exponential part of the time dependence $e^{-m\tau}$, $m=1,2,\ldots$.}
\be\label{WDF_basis}
\begin{aligned}
\mathcal{W}^{(l)}_{DF}&=\sum_{n=0}^{l-1}\tau^n \tilde h^{(l)}_n(\sigma)\,,\\
\tilde h^{(l)}_n(\sigma)&=\sum_{s,r,m_i} c^{\prime\pm}_{s,m_1,\ldots,m_r} e^{k \sigma}\sigma^s H_{m_1,\ldots,m_r}(-e^{-2\sigma})\,,\quad k=\pm2,0\,,
\end{aligned}
\ee
and provide explicit expressions for $\tilde h^{(l)}_n$ up to $l=6$ loops. These are included in the computer-readable file \texttt{WDF1-6.m} accompanying the version of this article on the \texttt{arXiv}.

\section{The 2-gluon bound state contribution}
Let us start by reviewing what is known about $R_6$ up to second order in the expansion around collinear kinematics~\cite{Basso:2014koa}. The kinematical dependence enters through the conformal cross ratios $u_i$ of the cusp positions $x_j$ shown in figure \ref{fig:WLduality}, which we parametrize as
\be
\begin{aligned}
u_1&=\frac{x_{46}^2x_{13}^2}{x_{36}^2 x_{14}^2} =\frac{1}{2}\frac{e^{2 \sigma+\tau}  \text{sech}\tau}{1+e^{2 \sigma }+ 2\,e^{\sigma -\tau } \cos \phi +e^{-2 \tau }}\,,\\
u_2&=\frac{x_{15}^2 x_{24}^2}{x_{14}^2 x_{25}^2}= \frac{1}{2} e^{-\tau } \text{sech}\tau  \,,  \\
u_3&=\frac{x_{26}^2 x_{35}^2}{x_{25}^2x_{36}^2} = \frac{1}{1+e^{2 \sigma }+ 2\,e^{\sigma -\tau } \cos \phi +e^{-2 \tau }}  \,. \label{crossratios}
\end{aligned}
\ee
Around the collinear limit $\tau\to\infty$, the remainder function has an expansion of the form
\be
R_6=\log\mathcal{W}-\log\mathcal{W}_{BDS}\,,\quad\mathcal{W}=1+e^{-\tau}\mathcal{W}_{\text{twist-1}}+e^{-2\tau}\mathcal{W}_{\text{twist-2}}+\mathcal{O}(e^{-3\tau})\,,
\ee
where $\mathcal{W}_{BDS}$ is a function known explicitly to all loops,\cite{Basso:2013aha} and 
\be\label{calW_excitations}
\begin{aligned}
\mathcal{W}_\text{twist-1}&=\cos\phi \mathcal{W}_F\,,\\
\mathcal{W}_\text{twist-2}&= \big(\mathcal{W}_{\phi\phi}+\mathcal{W}_{\psi{\bar \psi}}+\mathcal{W}_{F\bar F} \big) + 2\cos(2\phi) \big( \mathcal{W}_{FF}+\mathcal{W}_{DF} \big)\, , 
\end{aligned}
\ee
are the contributions of the flux tube excitations of the dual Wilson loop, consisting of gluons $F, \bar F$, fermions $\psi, \bar \psi$ and scalars $\phi$ of helicity $\pm 1, \pm 1/2$ and 0 respectively. In what follows we will restrict our attention to the contribution of the 2-gluon bound state $DF$,
\be\label{WDF}
\mathcal{W}_{DF}=\int_{-\infty}^{+\infty} \frac{du}{2\pi}  \mu(u) e^{-\gamma(u)\tau+i p(u)\sigma}.
\ee
In the last formula, the quantities $\gamma(u), p(u), \mu(u)$ are given to all loops in $g^2=\lambda/(4\pi)^2$ by
\be\label{measure_gluon}
\begin{aligned}
&\gamma(u)\equiv E(u)-2=4g\, \mathbb{Q} \cdot  \mathbb{M} \cdot  \kappa(u)\,,\quad\quad\quad p(u)=2u-4g\, \mathbb{Q} \cdot  \mathbb{M} \cdot  \tilde \kappa(u) \,,\\
&\mu(u) =\frac{\pi g^2 u\left(u^2+1\right)}{\sinh{(\pi u)}(x^{++}x^{--}-g^2)\sqrt{((x^{++})^2-g^2)((x^{--})^2-g^2)}}\times\\
&\exp{\bigg[\int\limits_{0}^{\infty}\frac{dt}{t}(J_{0}(2gt)-1)\frac{2e^{-t}\cos(ut)-J_{0}(2gt)-1}{e^{t}-1}\bigg]}e^{2\tilde \kappa(u) \cdot \mathbb{Q} \cdot  \mathbb{M} \cdot  \tilde\kappa(u)-2\kappa(u) \cdot \mathbb{Q} \cdot  \mathbb{M} \cdot  \kappa(u)}\, ,
\end{aligned}
\ee
where $\mathbb{Q}$ is a matrix with elements $\mathbb{Q}_{ij}=\delta_{ij}(-1)^{i+1}i$, $\mathbb{M}$ is related to another matrix $K$,
\be
\mathbb{M}\equiv (1+K)^{-1}=\sum_{n=0}^\infty (-K)^n\,,\quad \quad K_{ij}=2j(-1)^{j(i+1)} \int\limits_{0}^\infty \frac{dt}{t} \frac{J_i(2gt)J_j(2gt)}{e^t-1} \,,
\ee
$J_{i}$ is the $i$-th Bessel function of the first kind, and $\kappa, \tilde \kappa$ are vectors with elements
\be\label{kappa_vectors}
\begin{aligned}
\kappa_{j}(u) &\equiv  \int\limits_{0}^\infty \frac{dt}{t} \frac{J_j(2gt)(J_0(2gt)-\cos(ut)\left[e^{t/2}\right]^{(-1)^{j}-1})}{e^t-1} \\
\tilde\kappa_{j}(u) &\equiv  \int\limits_{0}^\infty \frac{dt}{t} (-1)^{j+1}\frac{J_j(2gt) \sin(u t) \left[e^{t/2}\right]^{(-1)^{(j+1)}-1}}{e^t-1}\,.
\end{aligned}
\ee
Finally, $x^{\pm\pm} = x(u\pm i)$ with $x(u) = \big(u+\sqrt{u^2-(2g)^2}\big)/2$.

\section{Method and Results}
Our main result is the proof that the integral (\ref{WDF}) evaluates to the basis (\ref{WDF_basis}) at any order $l$ in $g^2\ll 1$, and the derivation of explicit expressions for $\tilde h^{(l)}_n(\sigma)$ up to $l=6$. To this end, we employ the method developed in~\cite{Papathanasiou:2013uoa}, which consists of reducing the integral into a sum over residues, and using the technology of Z-sums~\cite{Moch:2001zr} in order to absorb the summation into the definition of HPLs,
\be
\quad\quad H_{m_1,\ldots,m_r}(x)=\sum_{n_1>n_2>\ldots >n_r\ge1}\frac{x^{n_1}}{n_1^{m_1}\ldots n_r^{m_r}}\,.
\ee

We have checked that our results for $\tilde h^{(l)}_n(\sigma)$ agree with the expansion of the full $R_6$ to 4 loops~\cite{Dixon:2014voa}, and also with the $\sigma\to -\infty$ limit given by $\mathcal{B}$ in p.26 of~\cite{Basso:2014koa}. For this we also need to compute $\mathcal{W}_{FF}$ to lowest order, which can be done along similar lines, see the appendix and also~\cite{Hatsuda:2014oza}. We close by writing a new prediction for part of $R^{(5)}_6$ (all HPLs have argument $-e^{-2\sigma}$, and $H_{i,(j,k)}=(H_{i,j,k}+H_{i,k,j})/2$),
\be\label{5loopprediction}
\begin{aligned}
\tilde h^{(5)}_4=&e^{2\sigma}\Bigg\{\tfrac{160}{3} H_{1,(1,3)}+16 H_{1,2,2}+32 H_{2,(1,2)}+32 H_{3,1,1}-128 H_{1,1,1,1,1}-2 H_5+(32 \sigma -8) H_{2,1,1}\\
&+\tfrac{16}{3}H_{1,4}+\tfrac{32}{3}(H_{2,3}+H_{3,2})+64 \sigma  H_{1,(1,2)}+\left(-32 \sigma ^2+64 \sigma -\tfrac{8 \pi ^2}{3}-48\right) H_{1,1,1}\\
&+\left(\tfrac{64}{3}-\tfrac{32 \sigma }{3}\right) H_{3,1}+\left(\tfrac{40}{3}-\tfrac{16 \sigma }{3}\right) H_{1,3}+\left(\tfrac{16 \sigma }{3}+8\right) H_{2,2}+(128 \sigma -64) H_{1,1,1,1}\\
&+ \left(\tfrac{1}{2}-2 \sigma \right)H_4+\left(-\tfrac{32 \sigma ^2}{3}+16 \sigma -\tfrac{8 \pi ^2}{9}\right) H_{1,2}+\left(-\tfrac{32 \sigma ^2}{3}+24 \sigma +\tfrac{8}{3}-\tfrac{8 \pi ^2}{9}\right) H_{2,1}\\
&+ \left(\tfrac{2 \sigma ^2}{3}-\tfrac{20 \sigma }{3}+\tfrac{\pi ^2}{18}+\tfrac{41}{4}\right)H_3+ \left(\tfrac{4 \sigma ^3}{3}-\tfrac{22 \sigma ^2}{3}+\tfrac{\pi ^2 \sigma }{3}+\tfrac{28 \sigma }{3}+\tfrac{16 \zeta (3)}{3}-\tfrac{11 \pi ^2}{18}+\tfrac{131}{24}\right)H_2\\
&+ \left(-\tfrac{2 \sigma ^4}{9}+\tfrac{16 \sigma ^3}{9}-\tfrac{\pi ^2 \sigma ^2}{9}-12 \sigma ^2+\tfrac{4 \pi ^2 \sigma }{9}+\tfrac{92 \sigma }{3}-\tfrac{16 \sigma  \zeta (3)}{3}+8 \zeta (3)-\pi ^2-\tfrac{47}{3}-\tfrac{7 \pi ^4}{1080}\right)H_1\\
&+ \left(\tfrac{32 \sigma ^3}{9}-16 \sigma ^2+\tfrac{8 \pi ^2 \sigma }{9}+48 \sigma +16 \zeta (3)-\tfrac{4 \pi ^2}{3}-\tfrac{92}{3}\right)H_{1,1}\Bigg\}+\tfrac{40}{3}(H_{1,3}+ H_{3,1})+\tfrac{8}{3}H_4\\
&+\left(-16 \sigma ^2+48 \sigma -\tfrac{4 \pi ^2}{3}-\tfrac{92}{3}\right) H_{1,1}+16 \sigma  (H_{1,2}+H_{2,1})+(64 \sigma -48) H_{1,1,1}+8 H_{2,2}\\
&+ \left(\tfrac{16 \sigma ^3}{9}-12 \sigma ^2+\tfrac{4 \pi ^2 \sigma }{9}+\tfrac{92 \sigma }{3}+8 \zeta (3)-\pi ^2-\tfrac{47}{3}\right)H_1-64 H_{1,1,1,1}+ \left(\tfrac{29}{3}-\tfrac{8 \sigma }{3}\right)H_3\\
&+\left(-\tfrac{16 \sigma ^2}{3}+12 \sigma -\tfrac{4 \pi ^2}{9}\right)H_2-\tfrac{\sigma ^4}{9}+\tfrac{14 \sigma ^3}{9}+\left(-\tfrac{28}{3}-\tfrac{\pi ^2}{18}\right) \sigma ^2+\sigma  \left(-\tfrac{8 \zeta (3)}{3}+\tfrac{7 \pi ^2}{18}+\tfrac{47}{3}\right)\\
&+\tfrac{20 \zeta (3)}{3}-\tfrac{7 \pi ^2}{9}-\tfrac{35}{48}-\tfrac{7 \pi ^4}{2160}+(\sigma\to-\sigma) \,.\nonumber
\end{aligned}
\ee
All $\tilde h^{(l)}_n(\sigma)$ up to $l=6$ may be found in the ancillary file accompanying this article on the \texttt{arXiv}.

\section*{Acknowledgments}

Based on an invited talk, presented at the Rencontres de Moriond, QCD and High Energy Interactions, 2014. We are grateful to B.~Basso, J.~Drummond and E.~Sokatchev for enlightening discussions, J.~Drummond for comments on the manuscript, and the organizers of Moriond for the hospitality and support. This work was supported by the French
National Agency for Research (ANR) under contract StrongInt (BLANC-SIMI-4-2011).

\appendix

\section*{Appendix: The contribution of two same-helicity gluons}
In this appendix, we calculate the contribution $\mathcal{W}_{FF}^{(l)}$ of two gluons with same helicity, to leading loop order $l=4$. As we mentioned in the main text, this was used to check our results against the collinear limit expansion of the full $R_6$ up to this order, given that $\mathcal{W}_{DF}$ and $\mathcal{W}_{FF}$ appear in a similar fashion in (\ref{calW_excitations}). We close the appendix with a discussion on how to generalize this calculation to higher loops.

The contribution in question is given by~\cite{Basso:2014koa}
\be
\mathcal{W}_{FF}^{(4)}= \int\int \frac{du\,dv}{(2\pi)^2} \,   \frac{\pi ^3 (u-v)   (\tanh (\pi  u)-\tanh (\pi  v))\,e^{2i(u+v)\sigma}}{2\left(
   u^2+\tfrac{1}{4}\right)^2 \left(v^2+\tfrac{1}{4}\right)^2 \cosh(\pi u) \cosh(\pi v)}  \, ,
\ee 
and by expanding the numerator and changing the integration variables, it is easy to see that it reduces to products of 1-fold integrals,
\be\label{WFF4_factorized}
\mathcal{W}_{FF}^{(4)}= \frac{\pi}{4}(I_1 I_4-I_2 I_3)\,,
\ee 
where
\be
I_k=\int du \frac{e^{2i u\sigma}}{\left(
   u^2+\tfrac{1}{4}\right)^2 \cosh(\pi u)}\times
   \begin{cases}1&\text{if } k=1\\
   u&\text{if } k=2\\
   \tanh (\pi  u)&\text{if } k=3\\
   u\tanh (\pi  u)&\text{if } k=4
   \end{cases}\,\,.
\ee
Even more conveniently, we only need to focus on two of them as the remaining two may be obtained by
\be
I_{k+1}=-\frac{i}{2}\frac{dI_k}{d\sigma}\,,\quad k=1,3\,.
\ee
The two integrals $I_1, I_3$ are the simplest representatives of the class considered in~\cite{Papathanasiou:2013uoa}, and may be computed by residues. In this manner, we obtain (as in (\ref{5loopprediction}), all HPLs have argument $-e^{-2\sigma}$, and one may equivalently write $H_m(x)=\text{Li}_m(x)$ )
\be
\begin{aligned}
\mathcal{W}_{FF}^{(4)}=&-\frac{i\pi}{8}\left(I_1\frac{dI_3}{d\sigma}-I_3\frac{dI_1}{d\sigma}\right)\\
=&2 \left[ \left(e^{-\sigma }+e^{\sigma }\right)^2H_3- \left(1+e^{-2 \sigma }\right) \left(\frac{4 \sigma ^3}{3}+\frac{\pi ^2 \sigma}{3} \right)\right]H_1-\left(e^{-\sigma }+e^{\sigma }\right)^2H_2^2\\
&- \left[\frac{8 \sigma ^3}{3}+4 \sigma ^2+\frac{2 \pi ^2 \sigma }{3}+\frac{\pi ^2}{3}+e^{-2 \sigma } \left(4 \sigma ^2+\frac{\pi ^2}{3}\right)\right]H_2\\
&- \left(4 \sigma ^2 +4 \sigma +\frac{\pi ^2}{3}+4 e^{-2 \sigma } \sigma\right)H_3-e^{-2 \sigma }\left(-\frac{4 \sigma ^4}{3}-\frac{2 \pi ^2 \sigma ^2}{3}+\frac{\pi ^4}{36}\right)\,.\\
\end{aligned}
\ee

Finally, let us comment on how we can proceed at higher loops. From the finite-coupling expressions for $\mathcal{W}_{FF}$, presented in section 3 and appendix C of~\cite{Basso:2014koa}, it is straightforward to prove that the phenomenon we observed in (\ref{WFF4_factorized}) persists to all orders in the weak coupling expansion. Namely, the 2-fold integral always reduces to a sum of products of 1-fold integrals. In particular, the only na\"ively inseparable $u,v$ dependence in the $\mathcal{W}_{FF}$ integrand may be recast as
\be
\begin{aligned}
\frac{1}{\Gamma(iu-iv)\Gamma(iv-iu)}&=\frac{(u-v)}{\pi}\sinh (\pi u-\pi v)\\
&=\frac{(u-v)}{\pi}\left[\sinh (\pi  u) \cosh (\pi  v)-\cosh (\pi  u) \sinh (\pi  v)\right],
\end{aligned}
\ee
and upon expanding we indeed end up with terms with factorized $u,v$ dependence.

In conclusion, the resulting 1-fold integrals may be evaluated with the method developed in~\cite{Papathanasiou:2013uoa}. We defer this task to a future publication.

\end{document}